\documentclass[]{aa}

\usepackage{graphicx}
\usepackage{psfig}
\usepackage{natbib}
\bibpunct{(}{)}{;}{a}{}{,} 

\newcommand{\Lya}{Ly-$\alpha$\ }
\newcommand{\Lyb}{Ly-$\beta$}
\newcommand{\CIV}{\ion{C}{iv}}

\newcommand{\CII}{\ion{C}{ii}}
\newcommand{\AlII}{\ion{Al}{ii}}
\newcommand{\SiIV}{\ion{Si}{iv}}

\newcommand{\SiII}{\ion{Si}{ii}}

\newcommand{\FeIII}{\ion{Fe}{iii}}
\newcommand{\FeII}{\ion{Fe}{ii}}

\newcommand{\NI}{\ion{N}{i}}
\newcommand{\HI}{\ion{H}{i}}
\newcommand{\OI}{\ion{O}{i}}
\newcommand{\SII}{\ion{S}{ii}}
\newcommand{\ArI}{\ion{Ar}{i}}
\newcommand{\NiII}{\ion{Ni}{ii}}
\newcommand{\cm}{cm$^{-2}$}
\newcommand{\kms}{km s$^{-1}$}
\newcommand{\lsim}{\raisebox{-5pt}{$\;\stackrel{\textstyle <}{\sim}\;$}}

\newcommand{\zem}{$z_{\rm em}$}
\newcommand{\zabs}{$z_{\rm abs}$}

\begin{document}

    \title{Hints of star formation at $z>6$: the
   chemical abundances of the DLA system in the QSO BRI
   1202-0725 ($z_{\rm abs}=4.383$)\thanks{Based on
   material collected with the European Southern
   Observatory Very Large Telescope operated on Cerro Paranal
          (Chile). Proposal 66.A-0594(A)}}

    \author{Valentina D'Odorico  \and 
     Paolo Molaro}	   

    \institute{INAF - Osservatorio Astronomico di Trieste, 
               via  G.B. Tiepolo, 11, I-34131 Trieste, Italy}

   \offprints{V. D'Odorico}

   \date{Received: July 23, 2003; accepted: October 30, 2003}

    \titlerunning{Hints of star formation at $z>6$}
 
    \authorrunning{}

\abstract{The Damped \Lya (DLA) absorber at redshift
      $z_{\rm abs}=4.383$ observed toward QSO BRI
      1202-0725 is studied by means of high resolution
      (FWHM~$\approx 7$ \kms) VLT-UVES spectra. We refine
      a previously determined Si abundance and derive
      with confidence abundances for C, N and O which are
      poorly known in DLAs. The [O/Fe] ratio is $\sim
      0.6$, but we cannot establish if iron is partially
      locked into dust grains. The [C/Fe]~$=0.08\pm
      0.13$ and [Si/C]~$=0.31\pm0.07$. [N/O] and [N/Si]
      are about $-1$, which is consistent with the majority of DLAs. 
      This value is much larger than the one observed for 
      the DLA toward QSO J~0307-4945 at $z_{\rm abs} = 4.466$. 
      The current interpretation of the bimodal distribution of N
      abundances in DLAs implies that large [N/$\alpha$]
      values correspond to relatively old
      systems. Adopting a scale time of 500 Myrs for full
      N production by intermediate mass stars, the onset
      of star formation in our DLA took place already at
      redshift $> 6$.  }

   \maketitle

%

\section{Introduction}

Observational evidence of early star formation is increasing 
thanks mainly to metal abundance analysis 
of $z>4$ QSOs found in the large sky surveys. 
Fe/Mg abundance ratios near or possibly even above solar were 
measured from the emission  lines of $z \approx 6$  QSO spectra 
\citep{freudling,maiolino}.   
Assuming this iron excess is a signature of SNIa production, 
a major episode of star formation must have taken place in 
these quasar  hosts  at  $z \ge 9$ to 
give birth to the progenitor stars.
\par\noindent
At the same time, results from the WMAP satellite favoured an early
reionization epoch of the intergalactic medium, possibly in the interval
$11 \lsim z \lsim 30$ \citep{bennett,kogut}, requiring the  
existence of Population III stars 
at very high redshifts \citep[e.g.][]{cen,ciardi03}. 
The nature of these early stars may be investigated
through the relic metals they left in the intergalactic 
medium, provided the gas was not reprocessed by 
the subsequent generation of stars. 
\par\noindent
A  remarkably precise way of measuring the elemental 
abundances  of the gas up to very high redshifts  
is represented by damped Lyman-$\alpha$ absorption 
(DLA) systems observed in the spectra of quasars. 
Indeed, their characteristic large \HI\ column density
($N($\HI$) \ge 2 \times 10^{20}$ \cm) assures that
ionization corrections can be neglected 
and very high resolution spectra now available to the
community allow excellent determinations of the
associated ionic column densities. 
    
In this paper we present VLT-UVES observations of the
QSO \object{BRI 1202-0725} \citep[\zem~=~4.69,][]{mcmahon}  
whose spectrum shows a DLA at \zabs=4.383, detected for
the first time by \citet{giallo94}, which is one of the few
highest-redshift DLA known \citep{song:cowie}.   
High resolution Keck observations of this QSO 
(FWHM = 6.6 \kms, $\lambda\lambda\,4900-9000$
\AA) were presented by \citet{lu96,lu98}.
Lower resolution spectra were obtained by
\citet{wampler96} and \citet{song:cowie} with NTT-EMMI
and Keck-ESI respectively.  
\citet{fontana96} did multi-band deep imaging of the 
quasar field and reported the detection of a galaxy 
at a separation of 2.2 arcsec from the QSO line of sight
that could be responsible for the DLA system. 
Follow-up spectroscopy of the object clarified instead 
that the galaxy was at the redshift of the QSO 
\citep{petitjean96,fontana98}.   
\par\noindent
Many studies were dedicated to this quasar which has
been thoroughly investigated  both in the optical  
and in the FIR and submillimeter bands, in particular for 
the presence of strong associated molecular emission lines 
\citep[e.g. ][ and references therein]{ohta}.

The paper structure is the following: Section 2 gives
details about the observations and the reduction process;
in Section 3 we present the analysis of the spectrum and the 
column densities derived for the metal ions associated with 
the DLA system. 
Section 4 is dedicated to the obtained abundance 
ratios, in particular of carbon, oxygen and silicon, and 
how they relate to observations in other DLAs. 
In Section 5 we focus on nitrogen and we
discuss its abundance in the framework of the present
production models.  

Throughout the paper we will adopt the usual 
cosmological model with H$_0 = 70$ km s$^{-1}$
Mpc$^{-1}$, $\Omega_{\rm m}=0.3$ and
$\Omega_{\Lambda}=0.7$. 
 
\section{Observations and data reduction}
 
In February and March 2001, high resolution spectra of
the QSO BRI 1202-0725 were obtained in service mode
with the UV and Visual Echelle Spectrograph
\citep[UVES,][]{dekker00} mounted on the Kueyen telescope of the
ESO VLT (Cerro Paranal, Chile). 
The journal of observations is reported in
Table~\ref{obs}. 

\begin{table}
\caption{Journal of observations}\label{obs}
\begin{tabular}{ccccc}
\hline
\hline
 date &   t$_{\rm exp}$ & No    & Coverage   &  S/N \\
 d/m/y &  (h)           &       &    \AA    &  \\
\hline
  3,5-7,14/2/2001 & 11 & 7  & 3050-3870 & \\
                 &       & 7  & 4780-5760; 5843-6814 & 45-20 \\
 18,19/3/2001    & 3 & 2  & 3740-4983 & \\ 
                 &       & 2  & 6714-8529; 8677-10000 & 30-15 \\
\hline
\\
\end{tabular}
\scriptsize{In the third column we give the number of
single frames} 
\end{table}

Spectra were taken in dichroic mode with a slit of
1.0'' and binning of 2x2 pixels. 
The overall resolution is $R \sim 43000$ (FWHM~$\simeq 7$
\kms). The wavelength coverage and the signal-to-noise ratios 
are reported in Table~\ref{obs}. 
The Lyman edge of the
DLA at \zabs~$\simeq 4.383$ absorbs completely  the spectrum
shortwards of $\lambda \simeq 5000$ \AA.

Data reduction was carried out by using the specific
UVES pipeline \citep[see][]{balle00} in the framework of  
the 99NOV version of the ESO reduction package, MIDAS.  
The final spectrum was obtained as a rebinned weighted sum 
of the single spectra output by the pipeline. 
%
The level of the continuum was determined by manually
selecting regions not affected by evident absorption and by
interpolating them with a spline function of 3rd degree.   

\section{Analysis} 

Atomic parameters for the identification and fitting of
the lines are taken from \citet{morton}. New oscillator
strengths are adopted for \SiII\ $\lambda\,1304$ and 1526
\citep{spitz:fitz} and \FeIII\ $\lambda\,1122$ (Morton 2002 
unpublished, J.~X. Prochaska private communication).   
Lines are fitted with Voigt profiles in the LYMAN
context of the MIDAS reduction package \citep{font:ball}.   

\begin{figure} 
\resizebox{\hsize}{!}{\includegraphics{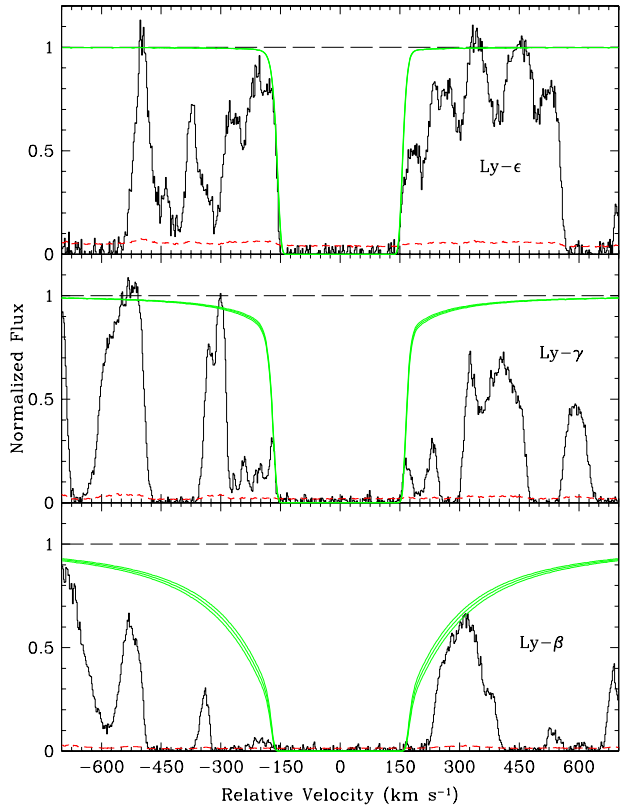}} 
\caption{Absorption lines of the Lyman series free from
strong blending (\Lya is not included to keep the velocity 
scale smaller). 
The thin solid line is the observed UVES
spectrum and the thin short-dashed line is the measured
standard deviation. The thick solid line is the fitted
profile obtained with one component at $z = 4.382724$
($v=0$ \kms) and $\log N($\HI$) = 20.55$, while the two 
thick short-dashed lines are drawn at the column
densities $\log N($\HI$)= 20.52$ and $20.58$}
\label{lymser} 
\end{figure}

Due to the lack of the fainter lines of the Lyman series
it was not possible to disentangle the velocity structure
of the \HI\ absorption lines. Thus, to determine the
total \HI\ column density we assumed a single component
at the average redshift of the strongest low ionization
absorption features ($z_{\rm Lyman} = 4.382724$).  
As a reference for the fit we took the blue wing of the
\Lya\ damped profile which is less contaminated by other
absorptions \citep[see Fig.~1 of][]{lu96} and we cared 
for having simultaneous agreement between the lines of the
Lyman series which were not strongly blended: 
\Lya, \Lyb, Ly-$\gamma$ and Ly-$\epsilon$.
We fitted the minimum and the maximum column densities  
consistent with the 4 velocity profiles and we took the mean 
of the two as the reference column density. 
The $1\,\sigma$ error is computed as 1/3 of the
difference between the  maximum and the minimum column
density (see Fig.~\ref{lymser}). 
The result is $\log N($\HI$) \simeq 20.55 \pm 0.03$ 
which slightly improves  the former  measure by
\citet{lu96}, $\log N($\HI$)=20.6 \pm 0.07$.

Outside the \Lya\ forest, we detected the following metal
absorption lines: \OI\ $\lambda\,1302$, 
\SiII\ $\lambda\lambda\, 1304,\ 1526$,  
\CII\ $\lambda\,1334$, 
\SiIV\ $\lambda\lambda\, 1393,\ 1402$, 
\CIV\ $\lambda\lambda\, 1548,\ 1550$ and 
\AlII\ $\lambda\,1670$ blended with telluric lines. 
Unfortunately, the absorption corresponding to \FeII\
$\lambda\,1608$ falls into one of the two gaps of the
spectrum. 

\begin{table}
\caption{Redshift, column density and Doppler parameters
of the components fitting the \SiII\ velocity profile}
\label{silicon}
\begin{tabular}{rccr}
\hline
\hline
$1\ldots$ &4.381440 & $12.74 \pm 0.04$  & $5.3 \pm 0.2$  \\
$2\ldots$ &4.381969 & $12.73 \pm 0.04$  & $7.0 \pm 0.2$  \\
$3\ldots$ &4.382644 & $13.44 \pm 0.05$  & $3.4 \pm 0.7$  \\
$4\ldots$ &4.382805 & $13.55 \pm 0.06$  & $3.5 \pm 0.7$  \\
$5\ldots$ &4.382973 & $13.19 \pm 0.19$  & $3.5 \pm 0.7$	 \\
$6\ldots$ &4.383096 & $13.66 \pm 0.09$  & $9 \pm 1.0$      \\
$7\ldots$ &4.383587 & $13.22 \pm 0.12$  & $7 \pm 1.0$   \\
$8\ldots$ &4.383901 & $13.68 \pm 0.06$  & $14 \pm 2.0$  \\
$9\ldots$ &4.384376 & $13.50 \pm 0.03$  & $9.3 \pm 0.7$  \\
$10\ldots$ &4.384813 & $12.57 \pm 0.06$  & $4.4 \pm 0.3$  \\
$11\ldots$ &4.385181 & $13.06 \pm 0.02$  & $6.2 \pm 0.1$
\end{tabular}
\end{table}

\subsection{Low-ionization lines} 

The common velocity profile of low ionization transitions was
well determined from the \SiII\ lines and fitted with 11
components (see Table~\ref{silicon}).
  
\OI\ $\lambda\,1302$ is saturated as in most DLAs. We
looked for fainter oxygen lines in the \Lya\ forest,
namely \OI\ $\lambda\, 936$, 948, 950, 971, 976, 988, 
1025, 1026 and 1039.  Although the majority of them was 
lost in the wealth of \Lya\ lines, we could set an upper
and a lower limit to the \OI\ total column density, $15.64 < \log
N($\OI$) < 15.83$, by constraining the strongest
components with the partially blended transitions \OI\
$\lambda\lambda\,971$ and 1039 and adopting the redshifts
and the Doppler parameters of \SiII. The latter assumption 
is based on the concordance among the redshifts and the Doppler 
parameters of the non-saturated components of \OI\ 
 with those of \SiII.  
The final column density is: $\log N($\OI$)=15.75 \pm 0.06$. 
\par\noindent
The central components of the \CII\ $\lambda\,1334$
absorption feature are also saturated. In order to get an
estimate of the C column density, we have followed the
same strategy adopted for oxygen relying on the partially
blended \CII\ $\lambda\, 1036$ absorption detected in
the \Lya\ forest. We assumed the parameters of \SiII\ which 
is a good tracer of \CII\ \citep{lev02}.   
We obtained a minimum and a maximum column density
consistent with the two profiles: $15.06 < \log N($\CII$)
< 15.2$, providing $\log N($\CII$)=15.13 \pm0.05$.

For the iron-peak elements we could only put an upper limit 
to the column density of \ion{Ni}{ii} 
$\lambda\,\lambda\, 1317,\ 1370$ \AA,
$\log($\ion{Ni}{ii}$) \lsim 12.5$.

\begin{figure} 
\resizebox{\hsize}{!}{\includegraphics{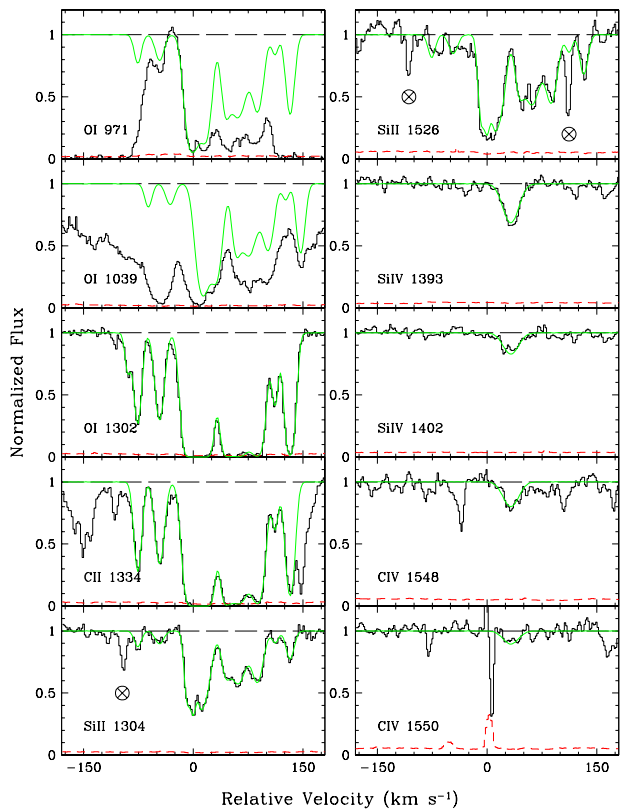}} 
\caption{Main metal absorption lines associated with the
DLA system at $z_{\rm abs}=4.383$. The solid thin line is
the normalized UVES spectrum and the solid thick line 
shows the fit of the velocity profile. The symbols
indicate the position of telluric lines}  
\label{metal438} 
\end{figure}

\begin{figure} 
\includegraphics[width=7cm,height=9cm,angle=-90]{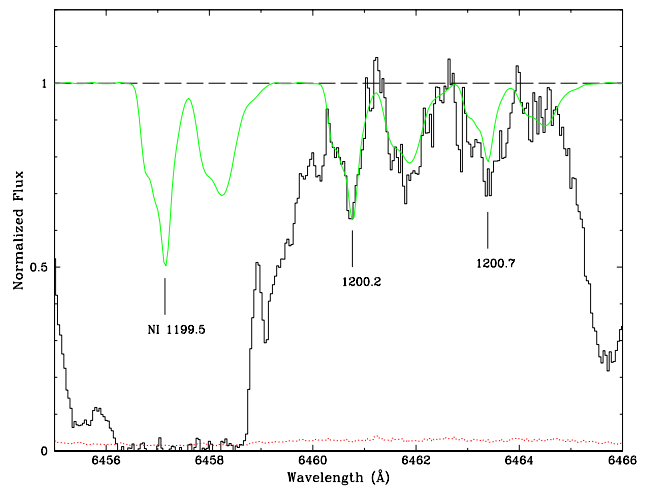} 
\caption{Nitrogen triplet \NI\ $\lambda\,1199.5,\
  1200.2,\ 1200.7$ \AA\ associated with the DLA system at
$z_{\rm abs} = 4.383$. The UVES spectrum is plotted as a function of 
  wavelength (thin solid line) and the thick line shows the fit of
  the velocity profile. The tick marks are drawn at $z=4.38297$ for
  the 3 transitions. } 
\label{azoto} 
\end{figure}

Inside the \Lya\ forest, we detected the lines
$\lambda\,1200.2$ and 1200.7 \AA\ of the \NI\ triplet
apparently free from blending (see Fig.~\ref{azoto}).
We fitted them with the central stronger components
detected in \SiII\ and obtained a column density which is
in agreement with the limit given by \citet{lu98}. 
The weaker \NI\ $\lambda\,1134$ triplet is lost in the 
forest, as are the absorption lines corresponding to
\SII\  $\lambda\,947$, $\lambda\lambda\lambda\, 1250,\
1253,\ 1259$ and \ArI\  $\lambda\lambda\,1048, 1066$.  

\begin{table}
\caption{Total column densities and abundances in  the
\zabs=4.383 DLA towards QSO 1202-0725}\label{colden} 
\begin{center}
\begin{tabular}{lllc}
\hline
\hline
 Ion & $\lambda_{rest}$  & log N (X)   & [X/H]$^a$    \\
\hline
 \HI  & 1215 & 20.55 $\pm$ 0.03 &  \\
 \CII & 1334 & 15.13 $\pm$ 0.05 & -2.01 $\pm$ 0.06 \\
 \CIV & 1548 & 12.86  $\pm$ 0.04 & \\
      & 1550 && \\
 \NI & 1200.2 & 13.81 $\pm$ 0.05 & -2.67 $\pm$ 0.06 \\
     & 1200.7 && \\
 \OI  & 1302 &  15.75 $\pm$ 0.06 & -1.54 $\pm$ 0.07 \\
      & 1039 & & \\
      & 971 && \\
\SiII & 1304 & 14.39 $\pm$ 0.05 & -1.70 $\pm$ 0.06 \\
      & 1526 && \\ 
\SiIV & 1393 & 12.71 $\pm$ 0.02 & \\
      & 1402 && \\
\FeIII & 1122 & 12.83 $\pm$ 0.09 & \\
\NiII  & 1317 & $<$~12.5 & $<$~-2.3 \\
       & 1370 & & \\
\hline
\end{tabular}
\end{center}
\scriptsize{$^a$   [X/H] = log (X/H) -- log (X/H)$_{\sun}$.
For N and O updated photospheric solar values given by
\citet{holweger} were used, for the rest of the 
elements we used the meteorite values given by
\citep{sol:abun}. 
}
\end{table}

\subsection{High-ionization lines}

We detected weak, single component \CIV\ and \SiIV\ doublets 
at a velocity corresponding to a minimum in the low ion 
absorption profile.  
At the same redshift we detected in the \Lya\ forest a
weak absorption line, possibly corresponding to \FeIII\
$\lambda\,1122$ whose fit gives: 
$\log N($\FeIII$) \lsim 12.83 \pm 0.09$.
The equivalent width of \CIV\ $\lambda\,1548$ is
$w = 0.03$ \AA, while that of \SiIV\ $\lambda\,1393$ is $w=0.04$ \AA. 
They are comparable with those of \Lya\ clouds with much smaller
\HI\ column densities \citep{songaila01,pettini03}. 
\par\noindent
The weakness and the central position of the high ions with 
respect to the low ions in this system are peculiar if compared 
with other DLAs. 
In the sample of 33 DLAs with high ions compiled by \citet{wp00a},
only BRI 1202-0725 and Q2237-0608 show \CIV\ spanning a velocity range  
smaller than that of low ions, they are also the only objects in the sample
with redshift larger than 4. 
A possible dependence on redshift of the relative kinematics is
also drawn from the result by \citet{ledoux98} who found a trend of 
decreasing \CIV\ to \OI\ velocity broadening ratio with increasing 
redshift for a sample of 26 DLA systems. 
On the other hand, the DLA at \zabs~$=4.466$ in the spectrum of BR
J0307-4945 shows \CIV\ and \SiIV\ more extended than the
low ions although not very strong \citep{mirka01}. 
The kinematics observed in BRI 1202-0725 and Q2237-0608
do not seem to be compatible either with the model of a
rotating disk with infalling material \citep{wp00b} or
with rotating protogalactic clumps \citep{haehnelt98}.  
We speculate that at large redshifts DLAs could probe objects
kinematically less disturbed than those at lower redshifts.  
More systems are needed to confirm this trend with redshift 
and to draw a reliable scenario. 

\citet{lu96} explained the weakness of the high ion 
absorptions associated with this DLA as due to a decrease with 
redshift of the metal enrichment in galactic disk and
halo gas and/or to the decrease of the mean intensity of
the UV ionizing background.  
Since the gas giving rise to the low ions shows a considerable 
amount of carbon (see next section), if the slab of ionized 
material is somewhat associated with the bulk of DLA gas 
we cannot invoke a  deficit in carbon to explain the weakness of \CIV. 
On the other hand, the ratio of \SiIV\ over \CIV\ equivalent width is
greater than 1, which suggests an ionization by a softer UV spectrum  
than at lower redshifts, as expected from the evolution of the 
ionizing UV background \citep[e.g.][]{haehnelt01}.

\vskip 12pt

The detected transitions and the fits are shown in
Fig.~\ref{metal438},  the total column densities are
reported in Table~\ref{colden}.  
In general, we increased the precision of the measured
column densities with respect to \citet{lu96}, and
measured \OI\ and \CII\ for which they gave only
limits.

\section{Metallicity and relative abundances}

\citet{lu96} measured a \FeII\ column density $13.88\pm
0.11$, giving [Fe/H]\footnote{Using the customary
definition [X/Y] = log(X/Y) - log(X/Y)$_{\sun}$.}~$=-2.17\pm
0.13$,  when the revised photospheric solar abundances 
by \citet{holweger} are adopted. 
A more recent determination is reported by
\citet{song:cowie}, $\log N($\FeII$)=14.09\pm 0.2$, 
which gives [Fe/H]~$=-1.96 \pm 0.2$. 
In the following, we will adopt the weighted mean of the 
two determinations: $\log N($\FeII$) = 13.91 \pm 0.11$ and 
[Fe/H]~$=-2.09 \pm 0.12$. 
In the presence of dust, iron is partially locked in the 
grains and a better estimator of metallicity is the
non-refractory  element zinc. 
In the case of our absorption system the 
two transitions \ion{Zn}{ii} $\lambda\,2026,\ 2062$ \AA\ cannot 
be observed because they fall into the near-IR region.

The limit on nickel abundance, [Ni/Fe]~$< -0.21$, might 
strengthen the trend of decreasing nickel over iron ratio in low 
[Fe/H] DLAs, while the unweighted mean of all DLA measurements 
is slightly over solar \citep{pw02}. 
This observational behaviour has not yet been understood and the  
nickel depletion onto dust is unknown \citep[see also][]{mirka01}. 


We were able to measure the abundances of carbon, 
nitrogen and oxygen, a rare event in DLAs.  
The transitions  \OI\ and \CII\ are in fact almost always 
saturated, while the \NI\ lines are often blended 
in the \Lya\ forest.  

\subsection{Carbon}

In the survey by \citet{pw02} there are 14 lower
limits of C abundances, all obtained from the saturated
\CII\ $\lambda\, 1334$ \AA\ line. \citet{lev02} and
\citet{lopez02} claimed C measurements with reasonable
error.
They accounted for the saturation of the \CII\
lines through detailed modelling of the velocity
structures and measured [C/Zn]~$\approx 0.25$. 
Also in the sample of sub-DLA systems collected by
\citet{celine03}, where C measurements are easier,
the [C/Fe] ratios are solar or slightly over-solar.  
\par\noindent
We found [C/H]~$=-2.01 \pm 0.06$, which gives a carbon
over iron ratio consistent with solar, or under-solar if
iron is affected by dust depletion.    
\citet{molaro03} measured an even lower value,
[C/Fe]~$\simeq -0.6$, for a DLA with non-saturated C at
\zabs~$=5.8$ towards SDSS 1044-0125. 
More carbon measurements are needed at large redshifts, 
maybe using sub-DLAs, to verify if, at variance with
what is observed for the intergalactic medium
\citep{pettini03}, the C abundance decreases with
increasing redshift in denser absorbers. 
This would give interesting hints on the enrichment
history of the universe and on the escape fraction of
metals from collapsed or collapsing systems.

\subsection{Oxygen and silicon}

Silicon is commonly measured in DLAs, while only a dozen 
of systems with some information on the oxygen abundance are 
presently available.
\par\noindent
We measured: [O/H]~$=-1.54\pm 0.07$, [Si/H]~$=-1.70\pm 
0.06$ and [O/Fe]~$=0.55\pm 0.14$,
[Si/Fe]~$=0.39\pm0.12$.   
The above $\alpha$/iron abundance ratios are in line with
the average values observed in most DLAs
\citep{pw02,molaro03}.   
We cannot state whether the observed values are due
entirely to $\alpha$-element enhancement or if iron is
partially depleted on dust grains (also silicon is mildly
refractory). 
There are several reasons to favour the latter
hypothesis: the observed metal column density measured
by the non-refractory element zinc, and the abundance
ratio [Fe/Zn] in DLA systems are anticorrelated
\citep{hou01}, while the observed [Si/Fe] and [Zn/Fe]
DLA abundance ratios are correlated \citep{pw02}. These
trends are expected if iron is affected by dust
depletion. Once corrected for dust, the Si/Fe
enhancement at [Fe/H]~$\sim-2/-1.5$ is typically
[Si/Fe]~$\simeq 0.2$ instead of $\simeq 0.3$ as observed
\citep{vladilo02} and, considering 26 DLAs where both Si
and the non-refractory element Zn were measured, the 
average value is $<$[Si/Zn]$>=-0.07 \pm 0.2$, consistent
with moderate, if any, enhancement.
In the three systems for which O and Zn are measured 
no genuine enhancement can be claimed with certainty 
\citep{molaro00,lopez02,ledoux03}.

\vskip 12pt

To conclude the review of the detected elements, we found
nitrogen much below the other elemental abundances at 
[N/H]~$=-2.67 \pm 0.06$, in agreement with the upper
limit set by \citet{lu98}. The abundance ratios to the
observed $\alpha$-elements are: [N/Si]~$=-0.97\pm0.08$
and [N/O]~$=-1.13\pm0.09$. In the next section we will
discuss the outcomes of these results.   

\vskip 12pt

Finally, we note that up to now the abundance patterns  
measured in the highest redshift DLAs do not resemble   
either the chemical pattern observed in PopII stars 
\citep{cayrel03} or 
the theoretical yields of massive Pop III stars
\citep{umeda:nomoto,heger:woos}.      
 
\section{Nitrogen and the epoch of star formation}

The production of nitrogen at very low
metallicities is still a debated subject.
Theoretically, it is assessed that primary N is created
in the thermal pulses of AGB stars undergoing 
hot bottom burning, namely the intermediate-mass
stars (4 to 7 M$_{\odot}$).

DLAs proved to be very useful to measure N abundances at 
metallicities lower than starburst galaxies. They
populate a plateau at [N/$\alpha$]~$\approx -0.8$
extending over two orders of magnitude in oxygen
abundance. 
\begin{figure}
      \begin{center}
      \includegraphics[width=6.5cm,height=8.5cm,angle=-90]{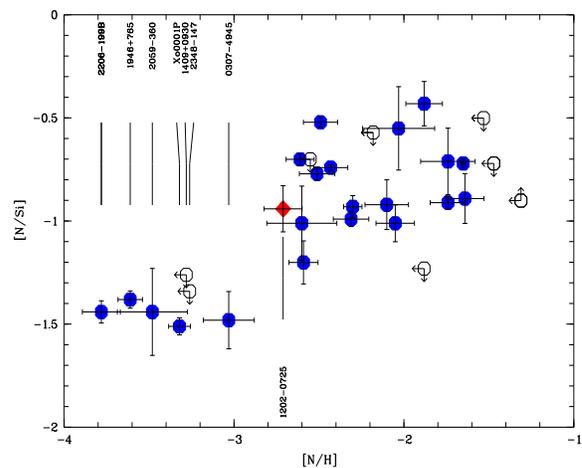} 
      \caption{[N/$\alpha$] ratio versus nitrogen
      abundance for the updated sample of DLAs. The empty circles are
      the upper and lower limits, our DLA is represented by the filled 
      diamond  (red if plotted in color). We have 
      labeled only the objects in the lower plateau and
      the studied QSO, BRI 1202-0725} 
      \label{nh}
      \end{center} 
\end{figure}
\begin{figure}
      \begin{center}
      \includegraphics[width=6.5cm,height=8.5cm,angle=-90]{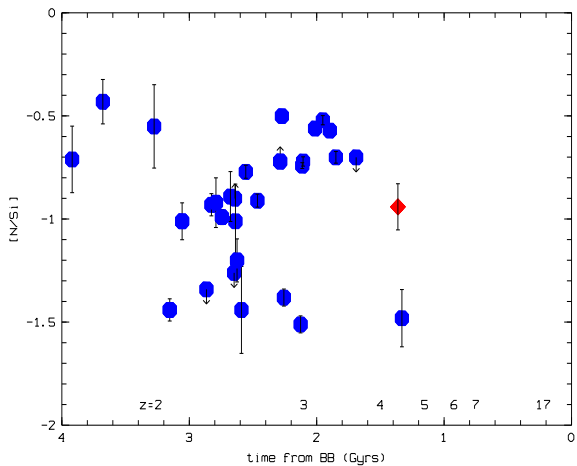}  
      \caption{[N/$\alpha$] ratio as a function of time from the Big
      Bang and of redshift. Our DLA (filled diamond) and the one in the
      spectrum of J~0307-4945 are the only two objects at
      $z > 4$}  
      \label{nsi-time}
      \end{center} 
\end{figure}
A few DLAs were found  
to form a possible second plateau with very little internal 
scatter at [N/$\alpha$] values about 0.7 dex below the main 
plateau \citep{proch02,centurio03}. 
Figure~\ref{nh} reports all the available measurements 
including the revised value for QSO 2206-199 by 
\citet{molaro03b}, the weighted mean in the low plateau 
is [N/Si]~=~$-1.446\pm 0.025$.  

Low [N/$\alpha$] values are expected in relatively young
objects where intermediate-mass stars did not yet have
time to evolve and to 
contribute to the nitrogen enrichment, while the short 
lived massive stars have already produced their oxygen. 
%
The low plateau has been interpreted as due to the effect of 
top-heavy initial mass function in these clouds
\citep{proch02} or to N synthesis by  the same massive
stars which produce the oxygen \citep{molaro03}. 
%
Within the latter scheme we expect to find high/low  
[N/Si] values in relatively old/young objects
respectively.  
Here, old and young refers to the characteristic
timescale of the main N production by intermediate-mass
stars, which can be of the order of 250-500 Myrs
\citep{henry00}, or longer if rotation plays an important 
role  \citep{mey:maed}.  
As a consequence, all 7 objects with low [N/Si]
abundance shown in Fig.~\ref{nsi-time} are young independently 
from their redshift, implying a continuos formation of
DLA absorbers at all epochs. 

The DLA system we have analised in this paper and the one 
along the line-of-sight to J~0307-4945 
\citep[$z_{\rm abs} = 4.466$, ][]{mirka01} are
the two highest redshift DLAs for which N has been
measured. 
The two systems have similar nitrogen abundances but
show remarkably different nitrogen over $\alpha$ abundances. 
The DLA in J~0307-4945 is positioned at the high N end of
the low plateau, while our DLA falls in the low N end of
the upper plateau (see Fig.\ref{nh}), as if it had just
made the jump, enriched by the N production of 
intermediate-mass stars. 

The highest redshift system is also the youngest,
according to the interpretation of the N abundance, but
it shows O abundance similar to our DLA. 
The observed pattern suggests that the system under study
is older but had a lower star formation rate (SFR)
generating a smaller amount of oxygen.
Then nitrogen might be considered a more reliable age  
indicator than oxygen, since the production of the latter  
is more dependent on the SFR.  
If, according to the chemical evolution models of
\citet{henry00} with low SFR, the time to attain the 
upper plateau level is of the order of 500
Myrs, the onset of star formation in the studied 
protogalaxy is placed at redshift higher than 6. 

\vskip 12pt

This measure provides additional evidence of early star 
formation. More refined theoretical models for the production 
of nitrogen and more nitrogen detections in very high redshift 
DLAs will allow us to get closer and closer to the epochs of 
formation of the very first stars predicted by the satellite 
WMAP at $11<z<30$ \citep{bennett}. 

\begin{acknowledgements}
We wish to thank our collaborators P. Bonifacio, 
M. Centuri\`on, C. P\`eroux and
G. Vladilo for valuable discussion on this topic. 
\end{acknowledgements}

\bibliographystyle{aa} 
\bibliography{aamnem99,myref} 

\end{document}